# The theorem on the magnetic field of rotating charged bodies

**Sergey G. Fedosin**

PO box 614088, Sviazeva str. 22-79, Perm, Perm Krai, Russia

E-mail: fedosin@hotmail.com

**Abstract:** The method of retarded potentials is used to derive the Biot-Savart law, taking into account the correction that describes the chaotic motion of charged particles in rectilinear currents. Then this method is used for circular currents and the following theorem is proved: The magnetic field on the rotation axis of an axisymmetric charged body or charge distribution has only one component directed along the rotation axis, and the magnetic field is expressed through the surface integral, which does not require integration over the azimuthal angle $\phi$. In the general case, for arbitrary charge distribution and for any location of the rotation axis, the magnetic field is expressed through the volume integral, in which the integrand does not depend on the angle $\phi$. The obtained simple formulas in cylindrical and spherical coordinates allow us to quickly find the external and central magnetic field of rotating bodies on the rotation axis.

## 1. Introduction

In the general case, stationary motion of a charged particle consists of rectilinear motion at a constant velocity and rotational motion at a constant angular velocity. Each of these motions in its own way leads to appearance of the corresponding magnetic field vector, so that the total magnetic field of a particle can be found by adding these two magnetic field vectors. If we consider stationary motion of a set of particles or motion of a charged body, then the total magnetic field of the system can be found based on the superposition principle as the sum of the magnetic field vectors of individual particles.

The result of flow of a rectilinear current of charged particles has been studied quite well, and for this case there is an experimentally derived Biot-Savart law, which can be written in a simplified form as follows [1]:

$$\mathbf{B}(1) = \frac{\mu_0}{4\pi} \int_2 \frac{[\mathbf{j} \times \mathbf{r}_{12}] dV}{r_{12}^3} , \qquad (1)$$

where $\mathbf{B}(1)$ is the magnetic field induction at a certain fixed point 1, calculated using the integral over the volume of area 2 occupied by the currents flowing in it; $\mu_0$ is the vacuum permeability; $\mathbf{j}$ is the electric current density vector inside area 2, depending on the coordinates, but not on the time; $\mathbf{r}_{12}$ is the vector from the point with the current inside area 2 to point 1; the quantity $\left[\mathbf{j} \times \mathbf{r}_{12}\right]$ is the vector product of $\mathbf{j}$ and $\mathbf{r}_{12}$.

There are various possible approaches, in which Equation (1) is found. As shown in [2], within the framework of the special theory of relativity, the magnetic field corresponding to (1) can be calculated as a consequence of the Lorentz transformations for the electromagnetic force, acting from one charged particle on another particle. Just as well, we can use the Lorentz transformation of the components of the electromagnetic tensor $F_{\mu\nu}$ from the moving reference frame $K'$, where there is only the electric field $\mathbf{E}'$, into the stationary reference frame $K$.

Indeed, by definition, $F_{\mu\nu} = \partial_\mu A_\nu - \partial_\nu A_\mu$, where $\partial_\mu = \dfrac{\partial}{\partial x^\mu} = \left(\dfrac{\partial}{c\,\partial t}, \dfrac{\partial}{\partial x}, \dfrac{\partial}{\partial y}, \dfrac{\partial}{\partial z}\right)$ is a four-gradient, $A_\mu = \left(\dfrac{\varphi}{c}, -\mathbf{A}\right)$ is a four-potential, expressed in terms of the scalar electric potential $\varphi$, the speed of light $c$ and the vector potential $\mathbf{A} = \left(A_x, A_y, A_z\right)$. Given that $A_\mu$ is a four-vector, and the same is true for $\partial_\mu$ in the special theory of relativity, the Lorentz transformations can be applied both to $A_\mu$ and to $\partial_\mu$. In this case, the components of any four-vector $a_\mu$ are transformed in the same way as the components of the four-dimensional quantity $x_\mu = \left(cdt, -x, -y, -z\right)$ that defines the location of a point in space and time. All this leads to the Lorentz transformations for the electromagnetic tensor components, so that in the coordinate notation we have

$$F_{\mu\nu} = \Lambda_\mu{}^\rho \Lambda_\nu{}^\sigma F'_{\rho\sigma},$$

where the four-dimensional quantities $\Lambda_\mu{}^\rho$ define the corresponding Lorentz transformation.

Since the nonzero tensor components equal $F_{01} = -F_{10} = \dfrac{E_x}{c}$, $F_{02} = -F_{20} = \dfrac{E_y}{c}$,

$F_{03} = -F_{30} = \dfrac{E_z}{c}$, $F_{32} = -F_{23} = B_x$, $F_{13} = -F_{31} = B_y$, $F_{21} = -F_{12} = B_z$, then for the

electromagnetic field components in $K$ we obtain the following:

$$E_x = E_x', \qquad E_y = \gamma\left(E_y' + V_0 B_z'\right), \qquad E_z = \gamma\left(E_z' - V_0 B_y'\right).$$

$$B_x = B_x', \qquad B_y = \gamma\left(B_y' - \frac{V_0 E_z'}{c^2}\right), \qquad B_z = \gamma\left(B_z' + \frac{V_0 E_y'}{c^2}\right). \tag{2}$$

In expressions (2) we set $\mathbf{B}' = \left(B_x', B_y', B_z'\right) = 0$, then the magnetic field in $K$ will equal

$\mathbf{B} = \left(0, -\dfrac{\gamma V_0 E_z'}{c^2}, \dfrac{\gamma V_0 E_y'}{c^2}\right)$, where $V_0$ is the velocity of motion of the reference frame $K'$ in $K$ along the axis $OX$, the Lorentz factor $\gamma = \dfrac{1}{\sqrt{1 - V_0^2/c^2}}$. Let us suppose that the electric field in

$K'$ arises from the static charge distribution with the constant charge density $\rho_{0q}$. This can be

written as follows

$$\mathbf{E}' = \frac{\rho_{0q}}{4\pi\varepsilon_0} \int \frac{\left(\mathbf{R}' - \mathbf{r}'\right) dV'}{\left|\mathbf{R}' - \mathbf{r}'\right|^3},$$

where $\mathbf{R}'$ is a vector from the distribution center to the observation point, $\mathbf{r}'$ is a vector

from the distribution center to an arbitrary point in the volume of the charge distribution, and

the integration is performed over the volume $V'$ of the charge distribution, which is fixed in

$K'$. Then the magnetic field in $K$, in view of the relations $c^2 \mu_0 \varepsilon_0 = 1$, $\mathbf{j} = \gamma \rho_{0q} \mathbf{V}_0$, can be

represented by the formula

$$\mathbf{B} = \frac{\mu_0}{4\pi} \int \frac{\left[\mathbf{j} \times \left(\mathbf{R}' - \mathbf{r}'\right)\right] dV'}{\left|\mathbf{R}' - \mathbf{r}'\right|^3}.$$

where primed quantities are specified in $K'$.

Let us suppose now that a certain volume $V$ is uniformly filled with a set of moving charge distributions, so that $V = \int dV'$, $dV = dV'$, and we need to find the total magnetic field outside the volume $V$. In this case, we can use the superposition principle for the magnetic fields. Instead of $\mathbf{R}' - \mathbf{r}'$ now we should use $\mathbf{r}_{12} = \mathbf{R} - \mathbf{r}_2$, where $\mathbf{R}$ is a vector from the center of the volume $V$ to observation point 1, $\mathbf{r}_2$ is a vector from the center of the volume $V$ to point 2 with the current density $\mathbf{j}$ in the volume $V$. This gives

$$\mathbf{B}(1) = \frac{\mu_0}{4\pi} \int_2 \frac{\left[ \mathbf{j} \times \mathbf{r}_{12} \right] dV}{r_{12}^3},$$

which coincides with (1). Expression (1) is also obtained from the general solution of the wave equation for the vector potential in case of constant currents [3].

Formula (1) is one of the main formulas in magnetostatics to calculate in the first approximation the magnetic field of constant distributed currents. As for the case of rotational motion of charged particles, this situation is more complex, since rotation is not described by an inertial reference frame. When a charged body rotates, the charged particles of this body also rotate, which leads to the current density $\mathbf{j}$, while the vector $\mathbf{j}$ is usually directed tangentially to the curves along which the charges rotate. The non-rectilinearity of the vector $\mathbf{j}$ inside the rotating charged body markedly affects the resulting magnetic field, and therefore a different formula is required instead of (1).

In [4-5] it was shown that the stationary magnetic field of axisymmetric rotating charge distribution, in principle, can be expressed in terms of the strength $\mathbf{E}$ and the scalar potential $\varphi$ of the electric field of this distribution. If the motion of charged particles is rectilinear, the magnetic field is expressed only in terms of $\mathbf{E}$ in full accordance with the Lorentz transformations for the components of the electromagnetic field tensor in inertial reference frames. This once again underlines the difference between rectilinear motion and rotational motion and the need for different formulas for the magnetic field, depending on the type of motion.

In this regard, our goal will be to derive relativistic formulas for magnetic fields arising from a system of constant currents and from a stationary rotating charge distribution. In contrast to the above approaches, the magnetic field will be calculated taking into account the intrinsic chaotic motion of charges. In the next section, we will briefly present the relativistic expression for the Biot-Savart law and estimate its accuracy, and in section 3 we will pass on to the analysis of rotational motion of charges and currents and to the proof of the theorem on the magnetic field for this case.

## 2. Rectilinear motion of charges

In order to estimate the accuracy of formula (1) for the magnetic field of stationary currents, it is necessary to proceed from the basis of the electromagnetic theory. As a starting point, we will use the method of retarded potentials [6-7], according to which the scalar and vector potentials outside of an arbitrarily moving charged point particle with the number $n$ are expressed by the formulas:

$$\varphi_n = \frac{q_n}{4\pi\varepsilon_0\left(\hat{R}_P - \hat{\mathbf{v}}\cdot\hat{\mathbf{R}}_P/c\right)}, \qquad \mathbf{A}_n = \frac{\mu_0 q_n \hat{\mathbf{v}}}{4\pi\left(\hat{R}_P - \hat{\mathbf{v}}\cdot\hat{\mathbf{R}}_P/c\right)} = \frac{\varphi_n \hat{\mathbf{v}}}{c^2}, \qquad (3)$$

where $q_n$ is the particle charge; $\varepsilon_0 = \dfrac{1}{\mu_0 c^2}$ is the vacuum permittivity; $\hat{\mathbf{v}}$ is the particle velocity at an early time point $\hat{t}$; $\hat{\mathbf{R}}_P$ is the vector from the charged particle to the point $P$, at which the potentials $\varphi_n$ and $\mathbf{A}_n$ are calculated; the vector $\hat{\mathbf{R}}_P$ has the length $\hat{R}_P$ and is calculated at an early time point $\hat{t}$; $c$ is the speed of light.

The early time point is defined by the formula:

$$\hat{t} = t - \frac{\hat{R}_P}{c}. \qquad (4)$$

The meaning of this equality lies in the fact that during the time $t - \hat{t}$ the electromagnetic action from the charge $q_n$ must travel the distance $\hat{R}_P$ at the speed $c$ to the point $P$ with the radius-vector $\mathbf{R} = (x, y, z)$ in order for the potentials $\varphi_n$ and $\mathbf{A}_n$ to appear at this point.

In the four-dimensional formalism of Minkowski spacetime, the characteristic of the electromagnetic field is the four-potential, which for the particle under consideration has the form:

$$A_\mu = \left( \frac{\varphi_n}{c}, -\mathbf{A}_n \right) = \frac{\varphi_n'}{c^2} u_\mu = \frac{\varphi_n'}{c^2} \left( \gamma_v c, -\gamma_v \mathbf{v} \right). \tag{5}$$

All the unprimed quantities in (5), including $\varphi_n$, $\mathbf{A}_n$, $\gamma_v$ and $\mathbf{v}$, are measured at the time $t$ in the reference frame $K$, in which the particle is moving. The subscript $\mu$ runs over the values $0, 1, 2, 3$, so that the four-potential component with the subscript $\mu = 0$ is related to the scalar potential: $A_0 = \frac{\varphi_n}{c}$.

In Cartesian coordinates $\mathbf{A}_n = \left( A_{nx}, A_{ny}, A_{nz} \right)$, therefore, according to (5), $A_1 = -A_{nx}$, $A_2 = -A_{ny}$, $A_3 = -A_{nz}$. The quantity $\varphi_n'$ is the scalar potential of the particle in the reference frame $K'$ associated with the particle; $u_\mu$ is the four-velocity of the particle. From (5) it follows that $\varphi_n = \gamma_v \varphi_n'$, where $\gamma_v = \frac{1}{\sqrt{1 - v^2/c^2}}$ is the Lorentz factor of the particle, besides, $\mathbf{A}_n = \frac{\varphi_n' \gamma_v \mathbf{v}}{c^2} = \frac{\varphi_n \mathbf{v}}{c^2}$, according to (3), at $\mathbf{v} = \hat{\mathbf{v}}$. The equality $\mathbf{v} = \hat{\mathbf{v}}$ means that the special theory of relativity in four-dimensional formalism is correct for point particles, either for inertial reference frames and in the absence of particle acceleration, or with the proviso that values in retarded time should be used. If these conditions are not met, then it is better to return to the original principles of the theory in the form of (3). For comparison, in [3] the formulas for retarded potentials are obtained based on the solutions of the wave equations for the potentials using the Lorentz gauge. Thus, it is shown that Maxwell equations and retarded potentials are consistent with each other.

Let us express the coordinates of the point $P$ in $K'$ in terms of the coordinates of this point in $K$ using the Poincaré transformations, taking into account that in $K$ the point $P$ is defined by the radius-vector $\mathbf{R}_n = (x, y, z)$:

$$\mathbf{R}_n' = \mathbf{R}_n - \mathbf{v}\, t\, \gamma_v + \frac{(\gamma_v - 1)\, \mathbf{v}\, (\mathbf{v} \cdot \mathbf{R}_n)}{v^2}. \tag{6}$$

It is assumed here that at $t = 0$ the origin of the coordinate systems in $K$ and $K'$ coincide, and the clock in $K'$ shows the time $t' = 0$.

When the charge $q_n$ is at the origin in $K'$, the potential $\varphi'_n$ at the point $P$ is expressed in terms of $R'_n$ and the coordinates in $K'$ in a standard way:

$$\varphi'_n = \frac{q_n}{4\pi\varepsilon_0 R'_n} = \frac{q_n}{4\pi\varepsilon_0 \sqrt{x'^2 + y'^2 + z'^2}}\,.$$

Let us substitute (6) here and express $\varphi'_n$ in terms of the coordinates $x, y, z$. To do this, it suffices to find the square of the length of the vector $\mathbf{R}'_n = (x', y', z')$ in terms of the vectors $\mathbf{R}_n$ and $\mathbf{v}$:

$$R'^2_n = R_n^2 + v^2 t^2 \gamma_v^2 + \frac{\gamma_v^2 (\mathbf{v}\cdot\mathbf{R}_n)^2}{c^2} - 2t\,\gamma_v^2 (\mathbf{v}\cdot\mathbf{R}_n)\,. \tag{7}$$

This allows us to express the scalar potential in $K'$ and the vector potential in $K$:

$$\varphi'_n = \frac{q_n}{4\pi\varepsilon_0 \sqrt{R_n^2 + v^2 t^2 \gamma_v^2 + \dfrac{\gamma_v^2 (\mathbf{v}\cdot\mathbf{R}_n)^2}{c^2} - 2t\,\gamma_v^2 (\mathbf{v}\cdot\mathbf{R}_n)}}\,.$$

$$\mathbf{A}_n = \frac{\varphi'_n \gamma_v \mathbf{v}}{c^2} = \frac{\mu_0 q_n \gamma_v \mathbf{v}}{4\pi \sqrt{R_n^2 + v^2 t^2 \gamma_v^2 + \dfrac{\gamma_v^2 (\mathbf{v}\cdot\mathbf{R}_n)^2}{c^2} - 2t\,\gamma_v^2 (\mathbf{v}\cdot\mathbf{R}_n)}}\,.$$

The time $t$ is present under the root sign, which leads to the dependence of the scalar and vector potentials on time. This is a consequence of the coordinates' transformation (6) and can be considered as a result of the terminal velocity of the electromagnetic effect propagation and the need to take into account its retardation in the Lienard-Wichert potentials (3). As a rule, the vector potential and the magnetic field of a moving charge are searched for at the time point $t = 0$, which allows us to simplify the expression for $\mathbf{A}_n$. In most cases, the magnitude of the velocity $\mathbf{v}$ is much less than the speed of light, which makes it possible to neglect the term $\dfrac{\gamma_v^2 (\mathbf{v}\cdot\mathbf{R}_n)^2}{c^2}$ in comparison with $R_n^2$. With this in mind, we have:

$$\mathbf{A}_n \approx \frac{\mu_0 \, q_n \, \gamma_v \, \mathbf{v}}{4\pi R_n} \, .$$

The distance $R_n$ from the observer's viewpoint in $K$ is the distance from the charge $q_n$ to the point $P$ where the vector potential $\mathbf{A}_n$ is sought.

Now let us assume that there is a set of closely spaced particles, forming compact spatial distribution, moving as a whole at the velocity $\mathbf{v}$. The total vector potential from the set of charged particles is obtained by integrating over the volume $V_d$ of the charge distribution, for which the relation $dq_n = \rho_q \, dV$ is used, where $\rho_q$ is the density of the moving charge, $dV$ denotes the differential of the moving volume, corresponding to the volume of one particle. Let us also consider the reference frame $K''$ associated with the center of this charge distribution. If we denote the vector from the charge distribution center to the point $P$ by $\mathbf{R}_{dP} = (x, y, z)$, and the vector from the charge distribution center to the charge $q_n$ by $\mathbf{r}_{dn} = (x_{dn}, y_{dn}, z_{dn})$, then we will obtain $\mathbf{R}_n = \mathbf{R}_{dP} - \mathbf{r}_{dn}$, $R_n = \sqrt{\left(x - x_{dn}\right)^2 + \left(y - y_{dn}\right)^2 + \left(z - z_{dn}\right)^2}$ . Taking all this into account, for the vector potential of the charge distribution we find:

$$\mathbf{A}_d \approx \frac{\mu_0}{4\pi} \int\limits_{V_d} \frac{\gamma_v \, \rho_q \, \mathbf{v} \, dV}{\sqrt{\left(x - x_{dn}\right)^2 + \left(y - y_{dn}\right)^2 + \left(z - z_{dn}\right)^2}} \, .$$

Let us now assume that there is a certain set of charge distributions, moving at different constant velocities in a sufficiently large space volume $V_\Sigma$, which is fixed relative to the reference frame $K$. In the limit of the currents continuously distributed over the volume, we can assume that $V_\Sigma$ is equal to the sum of all moving volumes of individual charge distributions: $V_\Sigma = \sum V_d$ .

In order to find the total vector potential, we need to sum up the vector potentials $\mathbf{A}_d$ of each charge distribution. Let us assume that $\mathbf{r}_{\Sigma d} = (x_{\Sigma d}, y_{\Sigma d}, z_{\Sigma d})$ is the vector from the center of the volume $V_\Sigma$ to the center of an arbitrary charge distribution, $\mathbf{R} = (x, y, z)$ is the vector from the center of the volume $V_\Sigma$ to the point $P$, so that $\mathbf{R}_n = \mathbf{R} - \mathbf{r}_{\Sigma d} - \mathbf{r}_{dn}$. Denoting $\mathbf{r}_{\Sigma d} + \mathbf{r}_{dn} = \mathbf{r}_\Sigma = (x_\Sigma, y_\Sigma, z_\Sigma)$, we obtain $\mathbf{R}_n = \mathbf{R} - \mathbf{r}_\Sigma$, $R_n = \sqrt{\left(x - x_\Sigma\right)^2 + \left(y - y_\Sigma\right)^2 + \left(z - z_\Sigma\right)^2}$ .

Now the sum of the vector potentials of charge distributions reduces to the integral over the fixed volume $V_\Sigma$:

$$\mathbf{A} \approx \frac{\mu_0}{4\pi} \int_{V_\Sigma} \frac{\gamma_v \, \rho_q \, \mathbf{v} \, dV}{\sqrt{(x-x_\Sigma)^2 + (y-y_\Sigma)^2 + (z-z_\Sigma)^2}} \, . \tag{8}$$

The quantities $x_\Sigma$, $y_\Sigma$ and $z_\Sigma$ denote the coordinates of the points of the volume $V_\Sigma$ and are specified relative to the center of the volume $V_\Sigma$.

Inside each individual charge distribution in the comoving reference frame $K''$, the particles have their own chaotic motion at a certain velocity $\mathbf{v}'$. In this case, using the vector rule of relativistic addition of velocities, for the absolute velocity $\mathbf{v}_K$ and the Lorentz factor $\gamma_K$ of an arbitrary particle in the reference frame $K$ we find:

$$\mathbf{v}_K = \frac{\mathbf{v}' + \frac{(\gamma_v - 1)(\mathbf{v}'\mathbf{v})}{v^2}\mathbf{v} + \gamma_v \mathbf{v}}{\gamma_v\left(1 + \frac{\mathbf{v}'\mathbf{v}}{c^2}\right)}, \qquad \gamma_K = \gamma'\gamma_v\left(1 + \frac{\mathbf{v}'\mathbf{v}}{c^2}\right), \tag{9}$$

where $\gamma_v = \dfrac{1}{\sqrt{1-v^2/c^2}}$ is the Lorentz factor for the velocity $\mathbf{v}$, $\gamma' = \dfrac{1}{\sqrt{1-v'^2/c^2}}$ is the Lorentz factor for the velocity $\mathbf{v}'$, and $\gamma_K = \dfrac{1}{\sqrt{1-v_K^2/c^2}}$ is the Lorentz factor for the velocity $\mathbf{v}_K$.

Taking into account the chaotic velocity $\mathbf{v}'$ in (9) leads to some change in the velocities of the particles in $K$, and in (8) the velocity $\mathbf{v}_K$ should be substituted instead of the velocity $\mathbf{v}$. But if we use the condition of chaotic motion of the charged particles in $K''$ and take into account a great number of these particles, we can simplify the problem by determining the average values for $\mathbf{v}_K$ and $\gamma_K$. With such averaging at a selected point inside the distribution of particles we should take a small volume adjacent to the point, containing a sufficient number of particles, and perform averaging over time and volume. In the first approximation, according to (9), we obtain $\bar{\mathbf{v}}_K \approx \mathbf{v}$, $\bar{\gamma}_K \approx \gamma' \gamma_v$. Then, instead of $\gamma_v$ in (8) we should substitute $\bar{\gamma}_K \approx \gamma' \gamma_v$ and take into account the definition for the current density $\mathbf{j} = \rho_q \mathbf{v}$ :

$$\mathbf{A} \approx \frac{\mu_0}{4\pi} \int_{V_\Sigma} \frac{\gamma' \gamma_v \, \mathbf{j} \, dV}{\sqrt{\left(x - x_\Sigma\right)^2 + \left(y - y_\Sigma\right)^2 + \left(z - z_\Sigma\right)^2}} \,. \tag{10}$$

In order to find the components of the vector potential $\mathbf{A}$, we need to take three integrals in (10), separately for each component of the current density $\mathbf{j} = \left(j_x, j_y, j_z\right)$. The quantities $\gamma'$, $\gamma_v$ and $\mathbf{j}$ are functions of the coordinates $x_\Sigma, y_\Sigma, z_\Sigma$ inside the fixed volume $V_\Sigma$. However, only the radius-vector $\mathbf{R} = (x, y, z)$, that defines the position of the point $P$ relative to the center of the volume $V_\Sigma$, depends on the coordinates $x, y, z$ of the point $P$. Therefore, when calculating the magnetic field by the formula $\mathbf{B} = \nabla \times \mathbf{A}$, the curl operation must only be applied to the quantity $\mathbf{R}$.

Introducing the curl under the integral sign in (10), and taking into account the equality $\sqrt{\left(x - x_\Sigma\right)^2 + \left(y - y_\Sigma\right)^2 + \left(z - z_\Sigma\right)^2} = \left|\mathbf{R} - \mathbf{r}_\Sigma\right|$ we obtain:

$$\mathbf{B} = \nabla \times \mathbf{A} \approx \frac{\mu_0}{4\pi} \int_{V} \frac{\gamma' \gamma_v \left[\mathbf{j} \times \left(\mathbf{R} - \mathbf{r}_\Sigma\right)\right] dV}{\left|\mathbf{R} - \mathbf{r}_\Sigma\right|^3} \,. \tag{11}$$

Expressions (11) and (1) coincide in appearance, taking into account that $\mathbf{R} - \mathbf{r}_\Sigma = \mathbf{r}_{12}$. Also, if in (11) we set $\gamma' = 1$, $\gamma_v = 1$, without taking into account the proper chaotic motion of the charged particles inside the matter and assuming that the velocity $\mathbf{v}$ of the charged particles is small in comparison with the speed of light $c$, then (11) turns into (1). Thus, the magnetic field in the Biot-Savart law (1) is determined with relative inaccuracy, equal in the order of magnitude to $v^2/c^2$.

### 3. Rotational motion of charges

Let the charge $q_n$ rotate about a certain axis at the constant angular velocity $\omega = \dfrac{d\phi}{dt}$ at a constant distance from this axis, equal to $\rho$. At the time point $t$, the linear rotational velocity is equal to

$$\mathbf{v}_r = \mathbf{v}_r(t) = \frac{d\mathbf{r}_q}{dt} = (-\omega\rho\sin\phi, \omega\rho\cos\phi, 0),$$

where the vector

$$\mathbf{r}_q = (\rho\cos\phi, \rho\sin\phi, z_d) = \left[\rho\cos(\omega t + \phi_0), \rho\sin(\omega t + \phi_0), z_d\right]$$

defines the position of the charge $q_n$ as a function of cylindrical coordinates $\rho, \phi, z_q$, and also as a function of time. The quantity $\phi_0$ denotes the initial phase, specifying the components $\mathbf{r}_q$ at $t = 0$.

Let us assume that the vector $\mathbf{R} = (x, y, z)$ connects the origin of coordinates and the point $P$, at which we need to find the magnetic field. Then the vector

$$\mathbf{R}_P = \mathbf{R} - \mathbf{r}_q = \left[x - \rho\cos(\omega t + \phi_0), y - \rho\sin(\omega t + \phi_0), z - z_d\right]$$

will be a vector from the charged particle to the point $P$ at the time point $t$. The length of this vector equals:

$$R_P = \sqrt{(x - \rho\cos\phi)^2 + (y - \rho\sin\phi)^2 + (z - z_d)^2} =$$
$$= \sqrt{R^2 + z_d^2 - 2zz_d + \rho^2 - 2\rho x\cos\phi - 2\rho y\sin\phi}.$$

According to (4), the early time point $\hat{t} = t - \dfrac{\hat{R}_P}{c}$ depends on the length $\hat{R}_P$ of the vector $\hat{\mathbf{R}}_P$, which is the vector $\mathbf{R}_P$, but taken at the early time point $\hat{t}$. Since $\phi = \omega t + \phi_0$, $\hat{\phi} = \omega\hat{t} + \phi_0$, then for the quantities at the early time point $\hat{t}$ we find:

$$\hat{\mathbf{r}}_q = (\rho\cos\hat{\phi}, \rho\sin\hat{\phi}, z_q) = \left[\rho\cos(\omega\hat{t} + \phi_0), \rho\sin(\omega\hat{t} + \phi_0), z_d\right],$$

$$\hat{\mathbf{v}}_r = \mathbf{v}_r(\hat{t}) = \frac{d\hat{\mathbf{r}}_q}{d\hat{t}} = (-\omega\rho\sin\hat{\phi}, \omega\rho\cos\hat{\phi}, 0), \tag{12}$$

$$\hat{\mathbf{R}}_P = \mathbf{R} - \hat{\mathbf{r}}_q = \left[ x - \rho \cos \hat{\phi}, \, y - \rho \sin \hat{\phi}, \, z - z_d \right],$$

$$\hat{R}_P = \sqrt{\left( x - \rho \cos \hat{\phi} \right)^2 + \left( y - \rho \sin \hat{\phi} \right)^2 + \left( z - z_d \right)^2} =$$
$$= \sqrt{R^2 + z_d^2 - 2zz_d + \rho^2 - 2\rho x \cos \hat{\phi} - 2\rho y \sin \hat{\phi}}. \tag{13}$$

Let us substitute into the Lienard-Wiechert formula (3) for the vector potential outside the moving point charged particle the rotational velocity $\hat{\mathbf{v}}_r$ instead of $\hat{\mathbf{v}}$, and take into account that $\hat{\mathbf{v}}_r \cdot \hat{\mathbf{R}}_P = \omega \rho y \cos \hat{\phi} - \omega \rho x \sin \hat{\phi}$:

$$\mathbf{A}_n = \frac{\mu_0 \, q_n}{4\pi} \left( \frac{\hat{\mathbf{v}}_r}{\hat{R}_P + \dfrac{\omega \rho x \sin \hat{\phi}}{c} - \dfrac{\omega \rho y \cos \hat{\phi}}{c}} \right)_n .$$

Now we need to sum up the vector potentials $\mathbf{A}_n$ of individual charged particles. Let there be some rotating body, containing a great number of closely spaced charged particles. These particles can also move chaotically at the velocity $\mathbf{v}'$ in the reference frame $K'$, which is fixed relative to the body.

For the case of rotation, the Lorentz factor $\gamma_v$ in (9) should be replaced with $\gamma_r$, and the velocity $\hat{\mathbf{v}}$ should be replaced with the rotational velocity $\hat{\mathbf{v}}_r$. After averaging of the Lorentz factor $\gamma_K$, its average value $\overline{\gamma}_K \approx \gamma' \gamma_r$ appears. The charge $dq_n$ of the rotating point particle in cylindrical coordinates can be expressed in terms of the invariant charge density $\rho_{0q}$, the Lorentz factors $\overline{\gamma}_K$ and $\gamma_r$, and in terms of the moving volume $dV$:

$$dq_n = \rho_q dV \approx \gamma' \rho_{0q} dV_\Sigma .$$

Here $\rho_q = \overline{\gamma}_K \rho_{0q} \approx \gamma' \gamma_r \rho_{0q}$ is the charge density of the charged matter moving with the Lorentz factor $\overline{\gamma}_K$; due to the Lorentz contraction we will obtain $dV = \dfrac{dV_\Sigma}{\gamma_r}$, where $dV_\Sigma = \rho \, d\rho \, d\phi \, dz_d$ is the volume element of a non-rotating body in cylindrical coordinates.

Taking this into account, the sum of the vector potentials of individual particles is transformed into an integral over the body's volume:

$$\mathbf{A} = \frac{\mu_0 \rho_{0q}}{4\pi} \int\limits_{v_{\Sigma}} \frac{\gamma' \hat{\mathbf{v}}_r \, \rho \, d\rho \, d\phi \, dz_d}{\hat{R}_P + \dfrac{\omega \rho x}{c} \sin \hat{\phi} - \dfrac{\omega \rho y}{c} \cos \hat{\phi}}.$$

If we take into account (12), then the vector potential $\mathbf{A}$ of the rotating body has two non-zero components:

$$A_x = -\frac{\mu_0 \omega \rho_{0q}}{4\pi} \int\limits_{v_{\Sigma}} \frac{\gamma' \sin \hat{\phi} \, \rho^2 \, d\rho \, d\phi \, dz_d}{\hat{R}_P + \dfrac{\omega \rho x}{c} \sin \hat{\phi} - \dfrac{\omega \rho y}{c} \cos \hat{\phi}}.$$

$$A_y = \frac{\mu_0 \omega \rho_{0q}}{4\pi} \int\limits_{v_{\Sigma}} \frac{\gamma' \cos \hat{\phi} \, \rho^2 \, d\rho \, d\phi \, dz_d}{\hat{R}_P + \dfrac{\omega \rho x}{c} \sin \hat{\phi} - \dfrac{\omega \rho y}{c} \cos \hat{\phi}}. \qquad (14)$$

The magnetic field is found by the formula $\mathbf{B} = \nabla \times \mathbf{A}$. Taking into account (14), we will calculate the component $B_z$ of the magnetic field:

$$B_z = \frac{\partial A_y}{\partial x} - \frac{\partial A_x}{\partial y} = \frac{\mu_0 \omega \rho_{0q}}{4\pi} \int\limits_{v_{\Sigma}} \frac{\partial}{\partial x} \left( \frac{\cos \hat{\phi}}{\hat{R}_P + \dfrac{\omega \rho x}{c} \sin \hat{\phi} - \dfrac{\omega \rho y}{c} \cos \hat{\phi}} \right) \gamma' \rho^2 \, d\rho \, d\phi \, dz_d +$$

$$+ \frac{\mu_0 \omega \rho_{0q}}{4\pi} \int\limits_{v_{\Sigma}} \frac{\partial}{\partial y} \left( \frac{\sin \hat{\phi}}{\hat{R}_P + \dfrac{\omega \rho x}{c} \sin \hat{\phi} - \dfrac{\omega \rho y}{c} \cos \hat{\phi}} \right) \gamma' \rho^2 \, d\rho \, d\phi \, dz_d.$$

Taking the partial derivatives, we find:

$$B_z = \frac{\mu_0 \omega \rho_{0q}}{4\pi} \int\limits_{V_z} \frac{\left(\begin{array}{l} \hat{R}_P \dfrac{\partial \cos\hat{\phi}}{\partial x} + \hat{R}_P \dfrac{\partial \sin\hat{\phi}}{\partial y} + \dfrac{\omega\rho x}{c}\sin\hat{\phi}\dfrac{\partial \cos\hat{\phi}}{\partial x} - \\[2mm] -\dfrac{\omega\rho y}{c}\cos\hat{\phi}\dfrac{\partial \sin\hat{\phi}}{\partial y} - \dfrac{\partial \hat{R}_P}{\partial x}\cos\hat{\phi} - \dfrac{\partial \hat{R}_P}{\partial y}\sin\hat{\phi} - \\[2mm] -\dfrac{\omega\rho x}{c}\cos\hat{\phi}\dfrac{\partial \sin\hat{\phi}}{\partial x} + \dfrac{\omega\rho y}{c}\sin\hat{\phi}\dfrac{\partial \cos\hat{\phi}}{\partial y} \end{array}\right)}{\left(\hat{R}_P + \dfrac{\omega\rho x}{c}\sin\hat{\phi} - \dfrac{\omega\rho y}{c}\cos\hat{\phi}\right)^2} \gamma' \rho^2\, d\rho\, d\phi\, dz_d.$$

(15)

From the relations $\phi = \omega t + \phi_0$, $\hat{\phi} = \omega \hat{t} + \phi_0$, $\hat{t} = t - \dfrac{\hat{R}_P}{c}$ it follows:

$$\hat{\phi} = \phi + \omega\left(\hat{t} - t\right) = \phi - \frac{\omega \hat{R}_P}{c},$$

$$\cos\hat{\phi} = \cos\phi \cos\frac{\omega \hat{R}_P}{c} + \sin\phi \sin\frac{\omega \hat{R}_P}{c}, \qquad \sin\hat{\phi} = \sin\phi \cos\frac{\omega \hat{R}_P}{c} - \cos\phi \sin\frac{\omega \hat{R}_P}{c}.$$

(16)

Therefore, for the partial derivatives we can write:

$$\frac{\partial \cos\hat{\phi}}{\partial x} = \frac{\omega}{c}\frac{\partial \hat{R}_P}{\partial x}\left(\sin\phi \cos\frac{\omega \hat{R}_P}{c} - \cos\phi \sin\frac{\omega \hat{R}_P}{c}\right) = \frac{\omega}{c}\frac{\partial \hat{R}_P}{\partial x}\sin\hat{\phi}.$$

$$\frac{\partial \cos\hat{\phi}}{\partial y} = \frac{\omega}{c}\frac{\partial \hat{R}_P}{\partial y}\left(\sin\phi \cos\frac{\omega \hat{R}_P}{c} - \cos\phi \sin\frac{\omega \hat{R}_P}{c}\right) = \frac{\omega}{c}\frac{\partial \hat{R}_P}{\partial y}\sin\hat{\phi}.$$

$$\frac{\partial \cos\hat{\phi}}{\partial z} = \frac{\omega}{c}\frac{\partial \hat{R}_P}{\partial z}\left(\sin\phi \cos\frac{\omega \hat{R}_P}{c} - \cos\phi \sin\frac{\omega \hat{R}_P}{c}\right) = \frac{\omega}{c}\frac{\partial \hat{R}_P}{\partial z}\sin\hat{\phi}.$$

$$\frac{\partial \sin\hat{\phi}}{\partial x} = -\frac{\omega}{c}\frac{\partial \hat{R}_P}{\partial x}\left(\sin\phi \sin\frac{\omega \hat{R}_P}{c} + \cos\phi \cos\frac{\omega \hat{R}_P}{c}\right) = -\frac{\omega}{c}\frac{\partial \hat{R}_P}{\partial x}\cos\hat{\phi}.$$

$$\frac{\partial \sin \hat{\phi}}{\partial y} = -\frac{\omega}{c} \frac{\partial \hat{R}_P}{\partial y} \left( \sin \phi \sin \frac{\omega \hat{R}_P}{c} + \cos \phi \cos \frac{\omega \hat{R}_P}{c} \right) = -\frac{\omega}{c} \frac{\partial \hat{R}_P}{\partial y} \cos \hat{\phi}.$$

$$\frac{\partial \sin \hat{\phi}}{\partial z} = -\frac{\omega}{c} \frac{\partial \hat{R}_P}{\partial z} \left( \sin \phi \sin \frac{\omega \hat{R}_P}{c} + \cos \phi \cos \frac{\omega \hat{R}_P}{c} \right) = -\frac{\omega}{c} \frac{\partial \hat{R}_P}{\partial z} \cos \hat{\phi}. \qquad (17)$$

Let us substitute (17) into (15):

$$B_z = \frac{\mu_0 \omega \rho_{0q}}{4\pi} \int\limits_{V_\Sigma} \frac{\left[ \begin{array}{c} \dfrac{\partial \hat{R}_P}{\partial x} \left( \dfrac{\omega \hat{R}_P}{c} \sin \hat{\phi} - \cos \hat{\phi} + \dfrac{\omega^2 \rho x}{c^2} \right) + \\[2mm] + \dfrac{\partial \hat{R}_P}{\partial y} \left( -\dfrac{\omega \hat{R}_P}{c} \cos \hat{\phi} - \sin \hat{\phi} + \dfrac{\omega^2 \rho y}{c^2} \right) \end{array} \right]}{\left( \hat{R}_P + \dfrac{\omega \rho x}{c} \sin \hat{\phi} - \dfrac{\omega \rho y}{c} \cos \hat{\phi} \right)^2} \gamma' \rho^2 \, d\rho \, d\phi \, dz_d.$$

From (13) and (17) we find:

$$\frac{\partial \hat{R}_P}{\partial x} = \frac{x - \rho \cos \hat{\phi}}{\hat{R}_P} - \frac{\omega \rho x}{c \hat{R}_P} \frac{\partial \hat{R}_P}{\partial x} \sin \hat{\phi} + \frac{\omega \rho y}{c \hat{R}_P} \frac{\partial \hat{R}_P}{\partial x} \cos \hat{\phi}.$$

$$\frac{\partial \hat{R}_P}{\partial y} = \frac{y - \rho \sin \hat{\phi}}{\hat{R}_P} - \frac{\omega \rho x}{c \hat{R}_P} \frac{\partial \hat{R}_P}{\partial y} \sin \hat{\phi} + \frac{\omega \rho y}{c \hat{R}_P} \frac{\partial \hat{R}_P}{\partial y} \cos \hat{\phi}.$$

$$\frac{\partial \hat{R}_P}{\partial z} = \frac{z - z_d}{\hat{R}_P} - \frac{\omega \rho x}{c \hat{R}_P} \frac{\partial \hat{R}_P}{\partial z} \sin \hat{\phi} + \frac{\omega \rho y}{c \hat{R}_P} \frac{\partial \hat{R}_P}{\partial z} \cos \hat{\phi}.$$

From here we express $\dfrac{\partial \hat{R}_P}{\partial x}$, $\dfrac{\partial \hat{R}_P}{\partial y}$ and $\dfrac{\partial \hat{R}_P}{\partial z}$, and then substitute $\dfrac{\partial \hat{R}_P}{\partial x}$ and $\dfrac{\partial \hat{R}_P}{\partial y}$ into the expression for $B_z$:

$$\frac{\partial \hat{R}_P}{\partial x} = \frac{x - \rho \cos \hat{\phi}}{\hat{R}_P + \dfrac{\omega \rho x}{c} \sin \hat{\phi} - \dfrac{\omega \rho y}{c} \cos \hat{\phi}}.$$

$$\frac{\partial \hat{R}_P}{\partial y} = \frac{y - \rho \sin \hat{\phi}}{\hat{R}_P + \dfrac{\omega \rho x}{c} \sin \hat{\phi} - \dfrac{\omega \rho y}{c} \cos \hat{\phi}} \, .$$

$$\frac{\partial \hat{R}_P}{\partial z} = \frac{z - z_d}{\hat{R}_P + \dfrac{\omega \rho x}{c} \sin \hat{\phi} - \dfrac{\omega \rho y}{c} \cos \hat{\phi}} \, . \tag{18}$$

$$B_z = \frac{\mu_0 \omega \rho_{0q}}{4\pi} \int\limits_{V_\Sigma} \frac{\left[\begin{array}{c} \rho - x\cos\hat{\phi} - y\sin\hat{\phi} + \dfrac{\omega \hat{R}_P x}{c}\sin\hat{\phi} - \dfrac{\omega \hat{R}_P y}{c}\cos\hat{\phi} + \\[2mm] + \dfrac{\omega^2 \rho \left(x^2 + y^2\right)}{c^2} - \dfrac{\omega^2 \rho^2 x}{c^2}\cos\hat{\phi} - \dfrac{\omega^2 \rho^2 y}{c^2}\sin\hat{\phi} \end{array}\right]}{\left(\hat{R}_P + \dfrac{\omega \rho x}{c}\sin\hat{\phi} - \dfrac{\omega \rho y}{c}\cos\hat{\phi}\right)^3} \gamma' \rho^2 \, d\rho \, d\phi \, dz_d . \tag{19}$$

Calculation of the components $B_x$ and $B_y$ turns out to be easier, since $A_z = 0$:

$$B_x = \frac{\partial A_z}{\partial y} - \frac{\partial A_y}{\partial z} = -\frac{\mu_0 \omega \rho_{0q}}{4\pi} \int\limits_{V_\Sigma} \frac{\partial}{\partial z}\left(\frac{\cos\hat{\phi}}{\hat{R}_P + \dfrac{\omega \rho x}{c}\sin\hat{\phi} - \dfrac{\omega \rho y}{c}\cos\hat{\phi}}\right) \gamma' \rho^2 \, d\rho \, d\phi \, dz_d .$$

$$B_y = \frac{\partial A_x}{\partial z} - \frac{\partial A_z}{\partial x} = -\frac{\mu_0 \omega \rho_{0q}}{4\pi} \int\limits_{V_\Sigma} \frac{\partial}{\partial z}\left(\frac{\sin\hat{\phi}}{\hat{R}_P + \dfrac{\omega \rho x}{c}\sin\hat{\phi} - \dfrac{\omega \rho y}{c}\cos\hat{\phi}}\right) \gamma' \rho^2 \, d\rho \, d\phi \, dz_d .$$

Using (17) and (18), we find:

$$B_x = -\frac{\mu_0 \omega \rho_{0q}}{4\pi} \int\limits_{V_\Sigma} \frac{\left(z - z_d\right)\left(\dfrac{\omega \hat{R}_P}{c}\sin\hat{\phi} - \cos\hat{\phi} + \dfrac{\omega^2 \rho x}{c^2}\right)}{\left(\hat{R}_P + \dfrac{\omega \rho x}{c}\sin\hat{\phi} - \dfrac{\omega \rho y}{c}\cos\hat{\phi}\right)^3} \gamma' \rho^2 \, d\rho \, d\phi \, dz_d .$$

$$B_y = \frac{\mu_0 \omega \rho_{0q}}{4\pi} \int\limits_{V_\Sigma} \frac{\left(z - z_d\right)\left(\dfrac{\omega \hat{R}_P}{c}\cos\hat{\phi} + \sin\hat{\phi} - \dfrac{\omega^2 \rho\, y}{c^2}\right)}{\left(\hat{R}_P + \dfrac{\omega \rho\, x}{c}\sin\hat{\phi} - \dfrac{\omega \rho\, y}{c}\cos\hat{\phi}\right)^3} \gamma'\,\rho^2\,d\rho\,d\phi\,d z_d \,. \qquad (20)$$

Let us place the origin of the coordinate system on the rotation axis, so that the axis $OZ$ would coincide with the rotation axis. Suppose now that the point $P$, where the magnetic field is sought, lies on the rotation axis. Then the vector from the origin of coordinates to the point $P$ will equal $\mathbf{R} = (0, 0, z)$, and, according to (13), will be $\hat{R}_P = \sqrt{\left(z - z_d\right)^2 + \rho^2}$. In view of (16), for the magnetic field components (20) at $x = y = 0$ we find:

$$B_x\left(OZ\right) =$$

$$= -\frac{\mu_0 \omega \rho_{0q}}{4\pi} \int\limits_{V_\Sigma} \frac{\left(z - z_d\right)\left(\begin{array}{l} \dfrac{\omega \hat{R}_P}{c}\sin\phi\cos\dfrac{\omega \hat{R}_P}{c} - \dfrac{\omega \hat{R}_P}{c}\cos\phi\sin\dfrac{\omega \hat{R}_P}{c} - \\[2mm] - \cos\phi\cos\dfrac{\omega \hat{R}_P}{c} - \sin\phi\sin\dfrac{\omega \hat{R}_P}{c} \end{array}\right)}{\left[\left(z - z_d\right)^2 + \rho^2\right]^{3/2}} \gamma'\,\rho^2\,d\rho\,d\phi\,d z_d .$$

$$B_y\left(OZ\right) =$$

$$= \frac{\mu_0 \omega \rho_{0q}}{4\pi} \int\limits_{V_\Sigma} \frac{\left(z - z_d\right)\left(\begin{array}{l} \dfrac{\omega \hat{R}_P}{c}\cos\phi\cos\dfrac{\omega \hat{R}_P}{c} + \dfrac{\omega \hat{R}_P}{c}\sin\phi\sin\dfrac{\omega \hat{R}_P}{c} + \\[2mm] + \sin\phi\cos\dfrac{\omega \hat{R}_P}{c} - \cos\phi\sin\dfrac{\omega \hat{R}_P}{c} \end{array}\right)}{\left[\left(z - z_d\right)^2 + \rho^2\right]^{3/2}} \gamma'\,\rho^2\,d\rho\,d\phi\,d z_d .$$

If the coordinate $z$ of the point $P$ is small, then the same could be said about the quantity $\hat{R}_P = \sqrt{\left(z - z_d\right)^2 + \rho^2}$. Then we can assume that $\dfrac{\omega \hat{R}_P}{c} << 1$ and expand $\sin\dfrac{\omega \hat{R}_P}{c}$ and $\cos\dfrac{\omega \hat{R}_P}{c}$ to the second-order terms. This simplifies the expressions for $B_x\left(OZ\right)$ and $B_y\left(OZ\right)$:

$$B_x(OZ) \approx$$

$$\approx \frac{\mu_0 \omega \rho_{0q}}{4\pi} \int_{V_\Sigma} \frac{(z-z_d)\left(\cos\phi + \dfrac{\omega^2 \hat{R}_P^2}{2c^2}\cos\phi + \dfrac{\omega^3 \hat{R}_P^3}{3c^3}\sin\phi - \dfrac{\omega^4 \hat{R}_P^4}{6c^4}\cos\phi\right)}{\left[(z-z_d)^2 + \rho^2\right]^{3/2}} \gamma' \rho^2 \, d\rho \, d\phi \, dz_d .$$

$$B_y(OZ) \approx$$

$$\approx \frac{\mu_0 \omega \rho_{0q}}{4\pi} \int_{V_\Sigma} \frac{(z-z_d)\left(\sin\phi + \dfrac{\omega^2 \hat{R}_P^2}{2c^2}\sin\phi - \dfrac{\omega^3 \hat{R}_P^3}{3c^3}\cos\phi - \dfrac{\omega^4 \hat{R}_P^4}{6c^4}\sin\phi\right)}{\left[(z-z_d)^2 + \rho^2\right]^{3/2}} \gamma' \rho^2 \, d\rho \, d\phi \, dz_d .$$

$$(21)$$

Suppose now that the rotating charged body is axisymmetric relative to the axis $OZ$. Then, in (21) integration over the coordinates $\rho$ and $z_d$ will be independent of the coordinate $\phi$. In this case, integration in (21) over the coordinate $\phi$ in the range from 0 to $2\pi$ will give zero, so that the components $B_x(OZ)$ and $B_y(OZ)$ on the rotation axis $OZ$ will become equal to zero.

Let us now consider the component $B_z$ in (19) on the axis $OZ$, where $x$ and $y$ are equal to zero, while $\hat{R}_P = \sqrt{(z-z_d)^2 + \rho^2}$:

$$B_z(OZ) = \frac{\mu_0 \omega \rho_{0q}}{4\pi} \int_{V_\Sigma} \frac{\gamma' \rho^3 \, d\rho \, d\phi \, dz_d}{\left[(z-z_d)^2 + \rho^2\right]^{3/2}} . \qquad (22)$$

We see that the integrand in formula (22) for $B_z$ does not contain any explicit function of the variable $\phi$, except for $d\phi$, in case of any choice of the rotation axis. In particular, the rotation axis can also be outside the rotating charged body.

For the body axisymmetric relative to the axis $OZ$, integration over the cylindrical variables $\rho$ and $z_d$ turns out to be independent of $\phi$. In this case, integration over the variable $\phi$ in the range from 0 to $2\pi$ will give just $2\pi$. Then, for (22) we obtain the following:

$$B_z(OZ) = \frac{\mu_0 \omega \rho_{0q}}{2} \int_{V_\Sigma} \frac{\gamma' \rho^3 \, d\rho \, dz_d}{\left[(z-z_d)^2 + \rho^2\right]^{3/2}} . \qquad (23)$$

Thus, we have proved the following theorem: The magnetic field on the rotation axis of an axisymmetric charged body or charge distribution has only one component directed along the rotation axis, and the magnetic field is expressed through the surface integral, which does not require integration over the azimuthal angle $\phi$. In the general case, for arbitrary charge distribution and for any location of the rotation axis, the magnetic field is expressed through the volume integral, in which the integrand does not depend on the angle $\phi$.

Expression (23) can be rewritten in spherical coordinates, taking into account that the volume element is $dV = \rho\, d\rho\, dz_d\, d\phi = r^2 dr \sin\theta\, d\theta\, d\phi$, and also $\rho = r\sin\theta$, $z_d = r\cos\theta$:

$$B_z\left(OZ\right) = \frac{\mu_0 \omega \rho_{0q}}{2} \int\limits_{V_z} \frac{\gamma' r^4 dr \sin^3\theta\, d\theta}{\left(z^2 - 2zr\cos\theta + r^2\right)^{3/2}}.$$ (24)

The absence of the need for integration over the variable $\phi$ in (23-24) simplifies calculation of the magnetic field on the rotation axis of the axisymmetric charge distribution.

## 4. Magnetic field on the cylinder's axis

Let us use (23) to calculate the magnetic field on the axis of a long solid cylinder, which has the length $L$, the radius $a$ and rotates at the angular velocity $\omega$. It is assumed that before the onset of rotation, this cylinder was uniformly charged over the entire volume with the charge density $\rho_{0q}$, and rotation does not lead to the charge shift due to the centrifugal force. For the solid cylinder we can also set the Lorentz factor $\gamma' = 1$ and thus neglect the proper chaotic motion of charged particles.

Placing the origin of the coordinate system at the center of the cylinder, from (23) we find:

$$B_z\left(OZ\right) = \frac{\mu_0 \omega \rho_{0q}}{2} \int\limits_{-L/2}^{L/2} \left(\int\limits_0^a \frac{\rho^3 d\rho}{\left[\left(z - z_d\right)^2 + \rho^2\right]^{3/2}}\right) dz_d.$$ (25)

In (25), the inner integral defines the field from the thin rotating charged disk with the radius $a$, located at the distance $z_d$ from the center of the cylinder, and the integral over the variable $z_d$ sums up the fields from all the thin disks located in the range from $-L/2$ to $L/2$ perpendicularly to the axis $OZ$.

Outside the cylinder at $z \geq L/2$ and at $z \leq -L/2$, the result of integration in (25) will be as follows:

$$\int_0^a \frac{\rho^3 \, d\rho}{\left[\left(z-z_d\right)^2 + \rho^2\right]^{3/2}} = \sqrt{\left(z-z_d\right)^2 + a^2} + \frac{\left(z-z_d\right)^2}{\sqrt{\left(z-z_d\right)^2 + a^2}} - 2\left(z-z_d\right).$$

$$B_z\left(z \geq L/2\right) = \frac{\mu_0 \, \omega \rho_{0q}}{2} \left[ \begin{array}{l} \left(z+\dfrac{L}{2}\right)\sqrt{\left(z+\dfrac{L}{2}\right)^2 + a^2} - \left(z-\dfrac{L}{2}\right)\sqrt{\left(z-\dfrac{L}{2}\right)^2 + a^2} + \\ + \left(z-\dfrac{L}{2}\right)^2 - \left(z+\dfrac{L}{2}\right)^2 \end{array} \right].$$

(26)

$$B_z\left(z \leq -L/2\right) = \frac{\mu_0 \, \omega \rho_{0q}}{2} \left[ \begin{array}{l} \left(\dfrac{L}{2}-z\right)\sqrt{\left(\dfrac{L}{2}-z\right)^2 + a^2} + \left(\dfrac{L}{2}+z\right)\sqrt{\left(\dfrac{L}{2}+z\right)^2 + a^2} + \\ + \left(\dfrac{L}{2}+z\right)^2 - \left(\dfrac{L}{2}-z\right)^2 \end{array} \right].$$

(27)

Formula (27) is obtained from (26) by replacing $z$ with $-z$. From (26-27) it follows that at the ends of the cylinder, where $z = L/2$ or $z = -L/2$, the magnetic field on the axis $OZ$ is the same and is equal in value to

$$B_z\left(z = L/2\right) = B_z\left(z = -L/2\right) = \frac{\mu_0 \, \omega \rho_{0q}}{2}\left(L\sqrt{L^2 + a^2} - L^2\right) \approx \frac{\mu_0 \, \omega \rho_{0q} a^2}{4}.$$ 
(28)

The approximate value in (28) corresponds to the case of a long cylinder, for which $L \gg a$.

Let us now find the magnetic field inside the cylinder. For example, let $0 \leq z \leq L/2$. We will cut the cylinder at such $z$ with a plane perpendicular to the axis $OZ$, and we will obtain two new cylinders, the lengths of which will be equal to $L/2 + z$ and $L/2 - z$, respectively. The magnetic field at the point $0 \leq z \leq L/2$ on the axis of the original cylinder will be equal to the sum of the magnetic fields at the ends of the two new cylinders. In (28) substituting instead of $L$ the lengths $L/2 + z$ and $L/2 - z$, respectively, and summing up the results, we find:

$$B_z\left(0 \leq z \leq L/2\right) = B_z\left(-L/2 \leq z \leq 0\right) = \frac{\mu_0\,\omega\,\rho_{0q}}{2}\left[\left(\frac{L}{2}+z\right)\sqrt{\left(\frac{L}{2}+z\right)^2+a^2}-\left(\frac{L}{2}+z\right)^2\right]+$$

$$+\frac{\mu_0\,\omega\,\rho_{0q}}{2}\left[\left(\frac{L}{2}-z\right)\sqrt{\left(\frac{L}{2}-z\right)^2+a^2}-\left(\frac{L}{2}-z\right)^2\right].$$

$$(29)$$

At $z = \pm L/2$ the magnetic field (29) coincides with the field (28) at the end of the cylinder. At $z = 0$ from (29) we find the following magnetic field at the center of the cylinder:

$$B_z\left(z=0\right) = \frac{\mu_0\,\omega\,\rho_{0q}}{2}\left(L\sqrt{\frac{L^2}{4}+a^2}-\frac{L^2}{2}\right) \approx \frac{\mu_0\,\omega\,\rho_{0q}\,a^2}{2}.$$

Comparison with (28) shows that the field at the center of a long cylinder is almost twice as large as the field on the axis $OZ$ at the ends of this cylinder. This difference is due to the increase in the field non-uniformity at the ends of the cylinder.

At large $z$, when $z \gg L/2$, it is possible to expand the roots in (26) to the second-order terms. As a result, we obtain:

$$B_z\left(z \gg L/2\right) \approx \frac{\mu_0\,\omega\,\rho_{0q}\,a^4 L}{8z^3}, \qquad (30)$$

so that the field on the cylinder's axis at large distances decreases in inverse proportion to the cube of the distance $z$ to the center of the cylinder.

### 4. Magnetic field on the ball's axis

Let us assume that there is a uniformly charged ball with the radius $a$, the invariant volume charge density $\rho_{0q}$, and rotating at the angular velocity $\omega$ about the axis $OZ$. For a solid ball, we can assume that the Lorentz factor of the charged particles is $\gamma' = 1$ in the reference frame rigidly associated with the ball.

We will first calculate the magnetic field outside the ball on the axis $OZ$. If the origin of the coordinate system is at the center of the ball, then for the field at $z \geq a$ and at $z \leq -a$, according to (24), we can write:

$$B_z(OZ) = \frac{\mu_0 \omega \rho_{0q}}{2} \int_0^a \left( \int_0^\pi \frac{\sin^3 \theta \, d\theta}{\left( z^2 - 2zr\cos\theta + r^2 \right)^{3/2}} \right) r^4 dr \, . \tag{31}$$

From (31) it follows:

$$\int_0^\pi \frac{\sin^3 \theta \, d\theta}{\left( z^2 - 2zr\cos\theta + r^2 \right)^{3/2}} = \frac{4}{3z^3} \, .$$

$$B_z(z \geq a) = \frac{2\mu_0 \omega \rho_{0q} a^5}{15 z^3}, \qquad B_z(z \leq -a) = -\frac{2\mu_0 \omega \rho_{0q} a^5}{15 z^3} \, . \tag{32}$$

In contrast to the field far from the rotating long cylinder in (30), the external field on the ball's rotation axis in (32) starts decreasing in inverse proportion to $z^3$ immediately outside the ball's the limits. At the pole of the ball at $z = a$ the magnetic field will equal $B_z(z = a) = \frac{2\mu_0 \omega \rho_{0q} a^2}{15}$ .

If we set $z = 0$ in (31), the field at the center of the ball is found:

$$B_z(z = 0) = \frac{\mu_0 \omega \rho_{0q}}{2} \int_0^a \left( \int_0^\pi \sin^3 \theta \, d\theta \right) r \, dr = \frac{\mu_0 \omega \rho_{0q} a^2}{3} \, .$$

The obtained estimates of the magnetic field correspond exactly to the values in [8].

## 5. Conclusion

Based on the Lienard-Wiechert expressions for retarded potentials (3), we derive the Biot-Savart law for the magnetic field in the form of (11). The peculiarity of the obtained expression is that it takes into account the proper chaotic motion of the charged particles inside the matter. This allows us to use (11) to analyze the electromagnetic field in the relativistic uniform system with freely moving charged particles.

The simplifications made in the derivation of (11) show that the Biot-Savart law has relative inaccuracy, equal in the order of magnitude to $v^2/c^2$. Here $c$ is the speed of light, $v$ denotes the average velocity of the charged particles in the current density $\mathbf{j} = \rho_q \mathbf{v}$, and $\rho_q$ is the charge density of the moving matter.

From rectilinear currents we pass on to stationary circular currents created by rotation of the charge distributions, and again use the Lienard-Wichert potentials. As a result, we arrive at the theorem on the magnetic field of rotating charged bodies. By exactly calculating the partial derivatives of the vector potential with respect to the coordinates, taking into account the retardation effect, we derive formulas (22-24) for the magnetic field on the rotation axis. According to the proven theorem, the magnetic field with any position of the rotation axis is defined by the integral over the charge distribution volume, while the integrand does not depend on the angular coordinate $\phi$. For axisymmetric bodies, the magnetic field on the rotation axis is always directed only along this axis, while the field does not depend on $\phi$, in accordance with the symmetry of these bodies.

In order to illustrate how the theorem works, we apply it first to a solid cylinder, and then to a ball, taking into account the fact that both bodies rotate and are uniformly charged over their volume. Formula (23) in cylindrical coordinates and formula (24) in spherical coordinates allow us to quickly and accurately determine the external magnetic field of rotating charged axisymmetric bodies, as well as the field at their center. In other cases, general formulas (21-22) should be used.

The proven theorem can also be used as an additional tool to determine electromagnetic fields when solving wave equations, which allows to determine the integration constants more precisely and to simplify gauging the obtained solutions for potentials and fields.